\newcommand{\gray}{\mbox{$\gamma$-ray}}

\newcommand{\pcmq}{\mbox{cm$^{-2}$}}
\newcommand{\psec}{\mbox{s$^{-1}$}}
\newcommand{\pkev}{\mbox{keV$^{-1}$}}
\newcommand{\funit}{\mbox{ph \pcmq \psec}}
\newcommand{\feunit}{\mbox{ph \pcmq \psec \pkev}}

\documentclass[a4paper]{ESLAB2005}
\usepackage{epsfig}

\begin{document}

\title{Prospects in space-based Gamma-Ray Astronomy}

\author{J. Kn\"odlseder
        (on behalf of a large community of European Gamma-Ray 
        Astronomers)}
\institute{Centre d'Etude Spatiale des Rayonnements, 
           9, avenue du Colonel-Roche,
           B.P. 4346,
	   31028 Toulouse Cedex 4,
	   France}

\maketitle

\begin{abstract}

At the uppermost part of the electromagnetic spectrum, observations of 
the gamma-ray sky reveal the most powerful sources and the most violent 
events in the Universe.
While at lower wavebands the observed emission is generally dominated 
by thermal processes, the gamma-ray sky provides us with a view on the 
non-thermal Universe, where particles are accelerated by still poorly 
understood mechanisms to extremely relativistic energies, and nuclear 
interactions, reactions, and decays are organising the basic elements 
of which our world is made of.
Cosmic accelerators and cosmic explosions are the major science 
themes that are addressed in this waveband.

With the unequalled INTEGRAL observatory, ESA has provided a unique 
tool to the astronomical community that has made Europe the world 
leader in the field of gamma-ray astronomy.
INTEGRAL provides an unprecedented survey of the soft gamma-ray sky, 
revealing hundreds of sources of different kinds, new classes of 
objects, extraordinary views of antimatter annihilation in our
Galaxy, and fingerprints of recent nucleosynthesis processes.

While INTEGRAL provides the longly awaited global overview over the soft 
gamma-ray sky, there is a growing need to perform deeper, more 
focused investigations of gamma-ray sources, comparable to the 
step that has been taken in X-rays by going from the ROSAT survey 
satellite to the more focused XMM-Newton observatory.
Technological advances in the past years in the domain of gamma-ray 
focusing using Laue diffraction techniques have paved the way towards 
a future European gamma-ray mission, that will outreach past missions 
by large factors in sensitivity and angular resolution.
Such a future {\em Gamma-Ray Imager} will allow to study particle 
acceleration processes and explosion physics in unprecedented depth, 
providing essential clues on the intimate nature of the most violent 
and most energetic processes in the Universe.

\end{abstract}

\section{Introduction}

Since the early days of space-science, the field of 
gamma-ray astronomy has played an important role in Europe.
As early as in 1975, ESA launched the COS-B satellite which provided 
the first extensive survey of the Galaxy in the energy range 50 MeV 
to 5 GeV.
Since then, European researchers have contributed in various manners 
to the development of the field, 
either by the design and exploitation of dedicated instruments, such 
as the 
French-Russian SIGMA telescope or the
Italian-Dutch BeppoSAX satellite, 
or by the participation in international collaborations, such as NASA's 
Compton Gamma-Ray Observatory (CGRO) or the recently launched Swift
mission.
These developments culminate today with the scientific exploitation of 
ESA's INTEGRAL observatory, a mission that actually places Europe as 
the world leader in the field.
With its unprecedented combination of good sensitivity and excellent 
imaging and spectroscopic capabilities, 
INTEGRAL provides a new and unique view of the soft gamma-ray sky, 
and will provide the setting of the field for the next decade.

In the process of defining ESA's space science program for the decade
2015--2025, it is legitimate to ask the question why our sky should be 
explored in gamma-rays, why gamma-ray astronomy should be performed in 
Europe, and what kind of instrument should follow the on-going 
INTEGRAL mission.
The aim of this paper is to suggest answers to these questions.
These answers have been collected from the European gamma-ray community 
during a prospects seminar that was held on March 18th, 2005 in Rome
(see the web-site 
{\tt http://www.cesr.fr/\mbox{$\sim$}jurgen/} {\tt rome2005/}).

\section{Why gamma-ray astronomy?}

As introductory remark, it is worth emphasising some unique features of 
gamma-ray astronomy:
the specific character of the emission processes,
the diversity of the emission sites, and
the penetrating nature of the emission.

First, the emission process that leads to gamma-rays is in general very 
specific, and as such, is rarely observable in other wavebands.
At gamma-ray energies, cosmic acceleration processes are dominant, 
while in the other wavebands thermal processes are generally at the 
origin of the emission.
For example, electrons accelerated to relativistic energies radiate 
gamma-ray photons of all energies through electromagnetic interactions 
with nuclei, photons, or intense magnetic fields.
Accelerated protons generate secondary particles through nuclear interactions, 
which may decay by emission of high-energy gamma-ray photons.
At gamma-ray energies, nuclear deexcitations lead to a manifold of 
line features, while in the other wavebands, it is the bound electrons 
that lead to atomic or molecular transition lines.
For example, the radioactive decay of tracer isotopes allows the 
study of nucleosynthesis processes that occur in the deep inner layers of 
stars.
The interaction of high-energy nuclei with the gas of the 
interstellar medium produces a wealth of excitation lines that probe 
the composition and energy spectrum of the interacting particles.
Finally, annihilation between electrons and positrons result in a unique 
signature at 511~keV that allows the study of antimatter in the Universe.

Second, the sites of gamma-ray emission in the Universe are very 
diverse, and reach from the nearby Sun up to the distant Gamma-Ray 
Bursts and the cosmic gamma-ray background radiation.
Cosmic acceleration takes place on all scales:
locally in solar flares, within our Galaxy (e.g.~in compact binaries, 
pulsars, supernova remnants), and also in distant objects (such as 
active galactic nuclei or gamma-ray bursts).
Cosmic explosions are another site of prominent gamma-ray emission.
They produce a wealth of radioactive isotopes, are potential sources 
of antimatter, and accelerate particles to relativistic energies.
Novae, supernovae and hypernovae are thus prime targets of gamma-ray 
astronomy.

Third, gamma-rays are highly penetrating, allowing the study of otherwise 
obscured regions.
Examples are regions of the galactic disk hidden by dense interstellar 
clouds, or the deeper, inner, zones of some celestial bodies, where the 
most fundamental emission processes are at work.
New classes of sources become visible in the gamma-ray domain, that 
are invisible otherwise.

In summary, gamma-ray astronomy provides a unique view of our Universe.
It unveils specific emission processes, a large diversity of emission 
sites, and probes deeply into the otherwise obscured high-energy 
engines of our Universe.
The gamma-ray Universe is the Universe of particle acceleration and 
nuclear physics, of cosmic explosions and non-thermal phenomena.
Exploring the gamma-ray sky means exploring this unique face of our 
world, the face of the evolving violent Universe.

\section{Cosmic accelerators}

\subsection{The link between accretion and ejection}

As a general rule, accretion in astrophysical systems is often 
accompanied by mass outflows, which in the high-energy domain take the 
form of (highly) relativistic jets.
Accreting objects are therefore powerful particle accelerators, that 
can manifest on the galactic scale as microquasars, or on the 
cosmological scale, as active galactic nuclei, such as Seyfert 
galaxies and Blazars.

Although the phenomenon is relatively widespread, the jet formation 
process is still poorly understood.
It is still unclear how the energy reservoir of an accreting system 
is transformed in an outflow of relativistic particles.
Jets are not always persistent but often transient phenomena, and it 
is still not known what triggers the sporadic outbursts in accreting 
systems.
Also, the collimation of the jets is poorly understood, and in 
general, the composition of the accelerated particle plasma is not 
known (electron-ion plasma, electron-positron pair plasma).
Finally, the radiation processes that occur in jets are not well 
established.

Observations in the gamma-ray domain are able to provide a number of 
clues to these questions.
Gamma-rays probe the innermost regions of the accreting systems 
that are not accessible in other wavebands, providing the closest 
view to the accelerating engine.
Time variability and polarisation studies provide important insights 
into the physical processes and the geometry that govern the 
acceleration site.
The accelerated plasma may reveal its nature through 
characteristic nuclear and/or annihilation line features which may 
help to settle the question about the nature of the accelerated plasma.

\subsection{The origin of galactic soft \gray\ emission}

\begin{figure}[!t]
  \begin{center}
    \epsfig{file=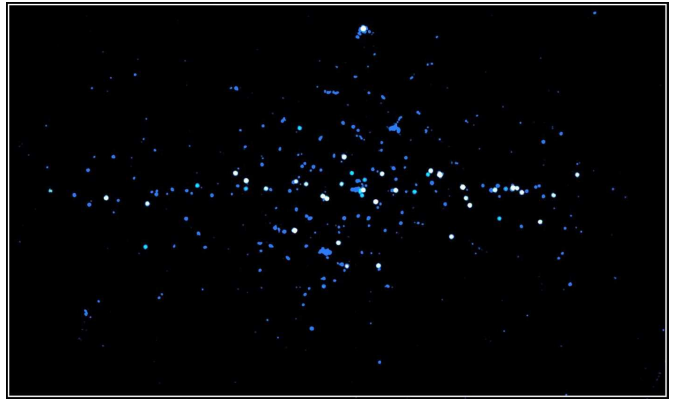, width=8.8cm}
  \end{center}
  \caption{The hard X-ray sky resolved into individual point sources 
    by the IBIS telescope aboard INTEGRAL
    (Lebrun et al.~2004).
    \label{fig:ibissky}}
\end{figure}

Since decades, the nature of the galactic hard X-ray\break ($>15$ keV)
emission has been
one of the most challenging mysteries in the field.
The INTEGRAL imager IBIS has now finally solved this puzzle.
At least $90\%$ of the emission has been resolved into point sources, 
settling the debate about the origin of the emission
(c.f.~Fig.~\ref{fig:ibissky}; \cite{lebrun04}).

At higher energies, say above $\sim300$ keV where the soft gamma-ray 
band starts, the situation is less clear.
In this domain, only a small part of the galactic emission has so far 
been resolved into point sources, and the nature of the bulk of the 
galactic emission is so far unexplained.
That a new kind of object or emission mechanism should be at work in 
this domain is already suggested by the change of the slope of 
the galactic emission spectrum.
While below $\sim300$ keV the spectrum can be explained by a 
superposition of Comptonisation spectra from individual point 
sources, the spectrum turns into a powerlaw above this energy, which 
is reminiscent of particle acceleration processes.
Identifying the source of this particle acceleration process, i.e. 
identifying the origin of the galactic soft gamma-ray emission, is 
one of the major goals of a future European gamma-ray mission.

One of the strategies to resolve this puzzle is to follow the 
successful road shown by INTEGRAL for the hard X-ray emission: trying 
to resolve the emission into individual point sources.
Indeed, a number of galactic sources show powerlaw spectra in the 
gamma-ray band, such as supernova remnants, like the Crab nebula, or 
some of the black-hole binary systems, like Cyg X-1
(\cite{mcconnell00}).
Searching for the hard powerlaw emission tails in these objects is 
therefore a key objective for a future gamma-ray mission.

\subsection{The origin of the soft \gray\ background}

After the achievements of XMM-Newton and Chandra, the origin of the
cosmic X-ray background (CXB) is now basically solved for energies close 
to a few keV. 
However, whilst the CXB is $\sim85\%$ and $80\%$ resolved in the\break
0.5--2~keV and 2--10~keV bands, respectively, it is only $\sim50\%$ 
resolved above $\sim8$ keV (\cite{worsley05}).
The situation is even worse in the soft gamma-ray band. 
Although about $20\%$ of the sources detected in the second IBIS 
catalogue are of extragalactic nature (\cite{bassani05}) they only 
account for $1\%$ of the background emission seen in the 20-100 keV band, 
i.e.~where the bulk of the soft \gray\ background energy density is found.

\begin{figure}[!t]
  \begin{center}
    \epsfig{file=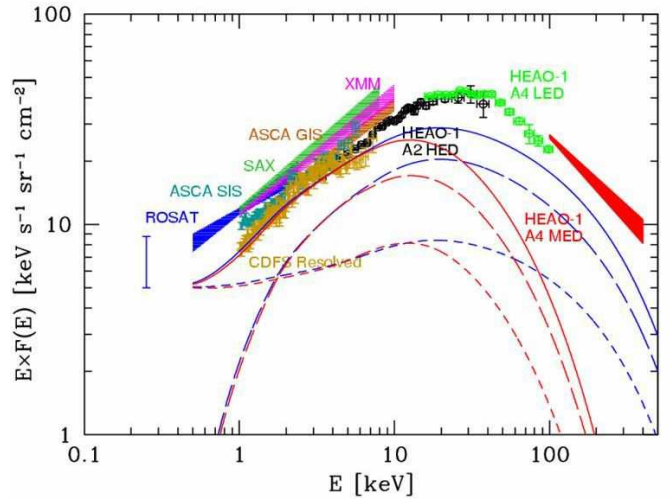, width=8.8cm}
  \end{center}
  \caption{The 0.25--400 keV cosmic X-/$\gamma-$ray background spectrum
     fitted with synthesis models; 
     measurements are from observations with different experiments as 
     labelled.
     Solid curves represent the integrated contribution of synthesis models: 
     blue corresponds to an AGN spectrum with cut-off at 400 keV while 
     and red to one with cut-off at 100 keV. 
     The short--dashed curves correspond to unabsorbed AGN 
     ($\log N_H < 22$ cm$^{-2}$), while the long--dashed curves correspond 
     to obscured Compton thin sources ($\log N_H$ in the range 22--24 cm$^{-2}$). 
     Figure from Comastri (2004).
    \label{fig:agn}}
\end{figure}

Looking from another point of view, synthesis models, which are well 
established and tested against observational results, can be used to 
evaluate the integrated AGN contribution to the soft \gray\ background. 
Unfortunately, they lack some key information at high energies: the 
absorption distribution is currently biased against low column densities 
due to the lack of soft gamma-ray surveys, no AGN luminosity function is 
available above 10 keV nor has the input spectral shape of the different 
classes of AGN been firmly established at high energies. 
Furthermore, the integrated AGN contribution changes as a function of 
model input parameters.
As an illustration, Fig.~\ref{fig:agn} shows how 
different results can be obtained by varying the power law energy cut-off.
A large region of  this parameter space is virtually unexplored because 
we currently lack information on large AGN samples.
Observations by BeppoSAX (\cite{risaliti02}; \cite{perola02}) of a 
handful of radio quiet sources, loosely locate this drop-off in the range 
30--300 keV; furthermore these measurements give evidence for a variable 
cut-off energy and suggest that it may increase with 
increasing photon index (\cite{perola02}). 
In radio loud sources the situation is even more complicated with some objects
showing a power law break and others no cut-off up to the MeV region. 
In a couple of low luminosity AGN no cut-off is present up to 300--500 keV. 
The overall picture suggests some link with the absence 
(low energy cut-off) or presence (high energy cut-off)  
of jets in the various AGN types sampled, but the data are still too scarce 
for a good understanding of the processes involved. 
One method to tackle this issue is to measure the soft gamma-ray SED 
(power law continuum plus high energy cut-off as well as hard tails if present) 
in a sizeable fraction of AGN in order to determine average shapes in 
individual classes and so  the nature of the radiation processes at the heart 
of all AGN.
This would provide at the same time information for soft \gray\ 
background synthesis models.
On the other hand, sensitive deep field observations should be able to 
resolve the soft \gray\ background into individual sources, allowing 
for the ultimate identification of the origin of the emission.

\subsection{Particle acceleration in extreme B-fields}

The strong magnetic fields that occur at the surface of neutron stars 
in combination with their fast rotation make them to powerful 
electrodynamic particle accelerators,\break which may manifest as pulsars 
to the observer.
Gamma-ray emitting pulsars can be divided into 3 classes:
spin-down powered pulsars, such as normal or millisecond pulsars,
accretion powered pulsars, occurring in low-mass or high-mass binary 
systems, and
magnetically powered pulsars, known as magentars.

Despite the longstanding efforts in understanding the physics of 
spin-down powered pulsars, the site of the\break gamma-ray production within 
the magnetosphere (outer gap or polar cap) and the physical process at 
action (synchrotron emission, curvature radiation, inverse Compton 
scattering) remain undetermined.
Although most of the pulsars are expected to reach their maximum 
luminosity in the MeV domain, the relatively weak photon fluxes have 
only allowed the study of a handful of objects so far.
Increasing the statistics will allow the study of the pulsar
lightcurves over a much broader energy range than today, providing 
crucial clues on the acceleration physics of these objects.

\begin{figure}[!t]
  \begin{center}
    \epsfig{file=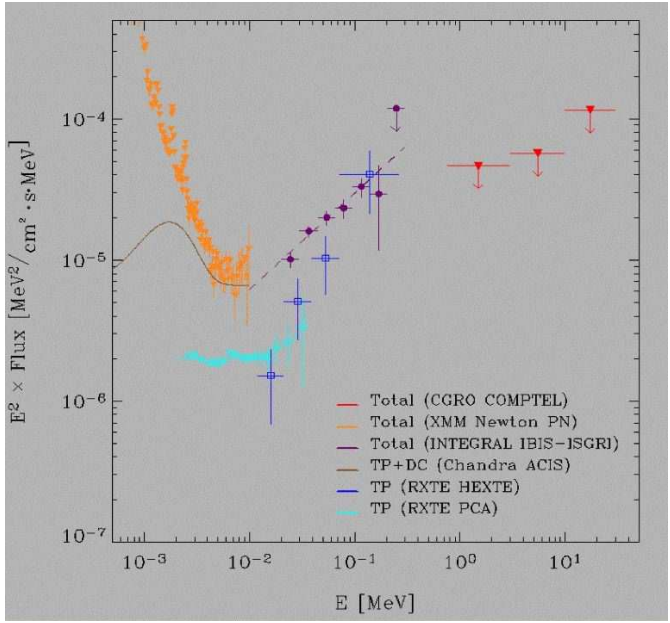, width=8.8cm}
  \end{center}
  \caption{High-energy spectral energy distribution of AXP 1E~1841-045
    (Kuiper et al.~2004).
    \label{fig:axp}}
\end{figure}

\begin{figure}[!t]
  \begin{center}
    \epsfig{file=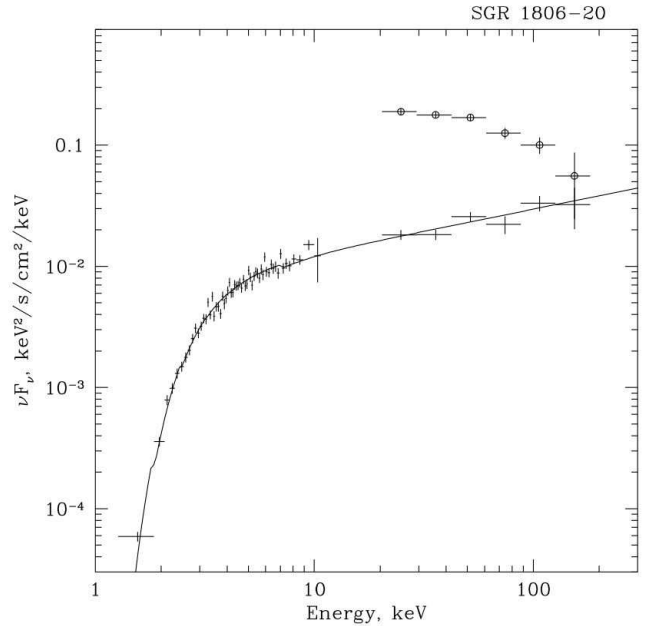, width=8.8cm}
  \end{center}
  \caption{The quiescent energy spectrum of SGR~1806-20 (lower 
    spectrum) and the summed spectrum of all detected bursts rescaled 
    down by a factor of 1000 (upper spectrum).
    From Molkov et al.~(2005).
    \label{fig:sgr}}
\end{figure}

Before the launch of INTEGRAL, the class of anomalous X-ray pulsars 
(AXPs), suggested to form a sub-class of the magnetar population, were 
believed to exhibit very soft X-ray spectra.
This picture, however, changed dramatically with the detection of 
AXPs in the soft gamma-ray band by INTEGRAL (\cite{kuiper04}).
In fact, above $\sim10$ keV a dramatic upturn is observed in the 
spectra which is expected to cumulate in the 100 keV -- 1 MeV domain
(c.f.~Fig.~\ref{fig:axp}).
The same is true for Soft Gamma-ray Repeaters (SGRs), as illustrated 
by the recent discovery of quiescent soft gamma-ray emission from 
SGR~1806-20 by INTEGRAL (c.f.~Fig.~\ref{fig:sgr}; \cite{molkov05}).
The process that gives rise to the observed gamma-ray emission in 
still unknown.
No high-energy cut-off has so far been observed in the spectra, yet 
upper limits in the MeV domain indicate that such a cut-off should be 
present.
Determining this cut-off may provide important insights in the 
physical nature of the emission process, and in particular, about the 
role of QED effects, such as photon splitting, in the extreme 
magnetic field that occur in such objects.
Strong polarisation is expected for the high-energy emission from 
these exotic objects, and polarisation measurements may reveal crucial 
to disentangle the nature of the emission process and the geometry of 
the emitting region.
Complementary measurements of cyclotron features in the spectra 
provide the most direct measure of the magnetic field strengths, 
complementing our knowledge of the physical parameters of the systems.

\section{Cosmic explosions}

\subsection{Understanding Type~Ia supernovae}

Although hundreds of Type~Ia supernovae are observed each year, and 
although their optical lightcurves and spectra are studied in great 
detail, the intimate nature of these events is still unknown.
Following common wisdom, Type~Ia supernovae are believed to arise in 
binary systems where matter is accreted from a normal star onto a 
white dwarf.
Once the white dwarf exceeds the Chandrasekhar mass limit a thermonuclear 
runaway occurs that leads to its incineration and disruption.
However, attempts to model the accretion process have so far failed to 
allow for sufficient mass accretion that would push the white dwarf 
over its stability limit (\cite{hillebrandt00}).
Even worse, there is no firm clue that Type~Ia progenitors are indeed 
binary systems composed of a white dwarf and a normal star.
Alternatively, the merging of two white dwarfs in a close binary 
system could also explain the observable features of Type~Ia events
(e.g.~\cite{livio03}).
Finally, the explosion mechanism of the white dwarf is only poorly 
understood, principally due to the impossibility to reliably model the 
nuclear flame propagation in such objects
(\cite{hillebrandt00}).

In view of all these uncertainties it seems more than surprising that 
Type~Ia are widely considered as standard candles.
In particular, it is this standard candle hypothesis that is the 
basis of one of the fundamental discoveries of the last decade: 
that the expansion of the Universe is currently accelerating
(\cite{riess98}).
Although empirical corrections to the observed optical lightcurves 
seem to allow for some kind of standardisation, there 
is increasing evidence that Type~Ia supernovae are not an homogeneous 
class of objects
(e.g.~\cite{mannucci05}).

\begin{figure}[!t]
  \begin{center}
    \epsfig{file=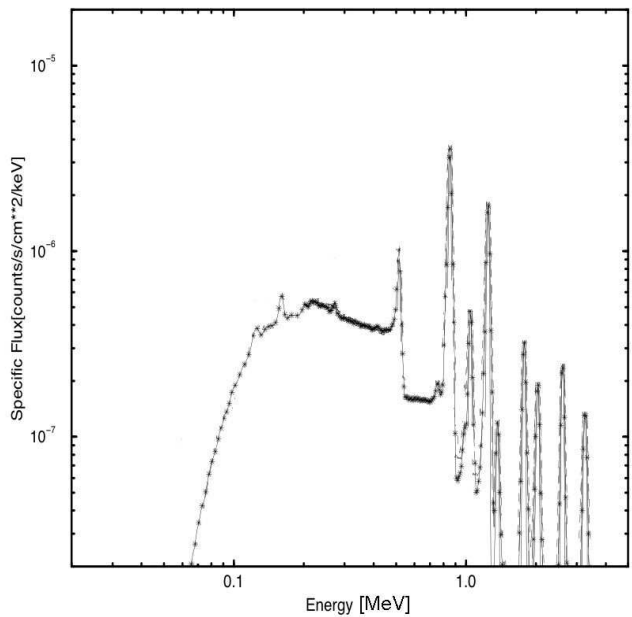, width=8.8cm}
  \end{center}
  \caption{Simulated gamma-ray spectrum of a Type~Ia supernova
    (Gomez-Gomar et al.~1998).
    \label{fig:snia}}
\end{figure}

Gamma-ray observation of Type~Ia supernovae provide a new and unique 
view of these events.
Nucleosynthetic products of the thermonuclear runaway lead to a rich 
spectrum of gamma-ray line and continuum emission that contains a 
wealth of information on the progenitor system, the explosion 
mechanism, the system configuration, and its evolution
(c.f.~Fig.~\ref{fig:snia}).
In particular, the radioactive decays of $^{56}$Ni and $^{56}$Co, 
which power the optical lightcurve which is so crucial for the 
cosmological interpretation of distant Type~Ia events, can be 
directly observed in the gamma-ray domain, allowing to pinpoint the 
underlying progenitor and explosion scenario.
The comparison of the gamma-ray to the optical lightcurve 
will provide direct information about energy recycling in the 
supernova envelope that will allow a physical (and not only empirical)
calibration of Type~Ia events as standard candles.

In addition to line intensities and lightcurves, the\break shapes of the 
gamma-ray lines hold important informations about the explosion 
dynamics and the matter stratification in the system.
Measuring the line shapes (and their time evolution) will allow to 
distinguish between the different explosion scenarios, ultimately 
revealing the mechanism that creates these most violent events in the 
Universe (\cite{gomez98}).

\subsection{Nova nucleosynthesis}

Classical novae are another site of explosive nucleosynthesis that is 
still only partially understood (see Hernanz et al., these proceedings).
Although observed elemental abundances in novae ejecta are relatively 
well matched by theoretical models, the observed amount of matter 
that is ejected substantially exceeds expectations.
How well do we really understand the physics of classical novae?

Radioactive isotopes that are produced during the nova explosion can 
serve as tracer elements to study these\break events.
Gamma-ray lines are expected from relatively long living isotopes, 
such as $^{7}$Be and $^{22}$Na, and from positron annihilation of 
$\beta^{+}$-decay positrons arising from the short living $^{13}$N and 
$^{18}$F isotopes.
Observation of the gamma-ray lines that arise from these isotopes
may improve our insight into the physical processes that govern the 
explosion.
In particular, they provide information on the composition of the white 
dwarf outer layers, the mixing of the envelop during the explosion, and 
the nucleosynthetic yields.
Observing a sizeable sample of galactic nova events in gamma-rays should 
considerably improve our understanding of the processes at work, and 
help to better understand the underlying physics.

\subsection{Understanding core-collapse explosions}

Gamma-ray line and continuum observations address some of the most 
fundamental questions of core-collapse supernovae: 
how and where the large neutrino fluxes couple to the stellar ejecta; 
how asymmetric the explosions are, including whether jets form; 
and what are quantitative nucleosynthesis yields from both static and 
explosive burning processes?

The ejected mass of $^{44}$Ti, which is produced in the innermost ejecta 
and fallback matter that experiences the alpha-rich freezeout of nuclear 
statistical equilibrium, can be measured to a precision of several 
percent in SN 1987A. 
Along with other isotopic yields already known, this will provide an 
unprecedented constraint on models of that event. 
$^{44}$Ti can also be measured and mapped, in angle and radial velocity, 
in several historical galactic supernova remnants. 
These measurements will help clarify the ejection dynamics, including 
how common jets initiated by the core collapse are.

Wide-field gamma-ray instruments have shown the\break global diffuse emission 
from long-lived isotopes $^{26}$Al and $^{60}$Fe, illustrating clearly 
ongoing galactic nucleosynthesis. 
A necessary complement to these are high-sensitivity measurements of the 
yields of these isotopes from individual supernovae. 
A future European gamma-ray mission should determine these yields, and map 
the line emission across several nearby supernova remnants, shedding further 
light on the ejection dynamics. 
It is also likely that the nucleosynthesis of these isotopes in hydrostatic 
burning phases will be revealed by observations of individual nearby massive 
stars with high mass-loss rates.

For rare nearby supernovae, within a few Mpc, we will be given a glimpse 
of nucleosynthesis and dynamics from short-lived isotopes $^{56}$Ni, and 
$^{57}$Ni, as was the case for SN 1987A in the LMC. 
In that event we saw that a few percent of the core radioactivity was 
somehow transported to low-optical depth regions, perhaps surprising 
mostly receding from us, but there could be quite some variety, especially 
if jets or other extensive mixing mechanisms are ubiquitous.

\subsection{Unveiling the origin of galactic positrons}

\begin{figure}[!t]
  \begin{center}
    \epsfig{file=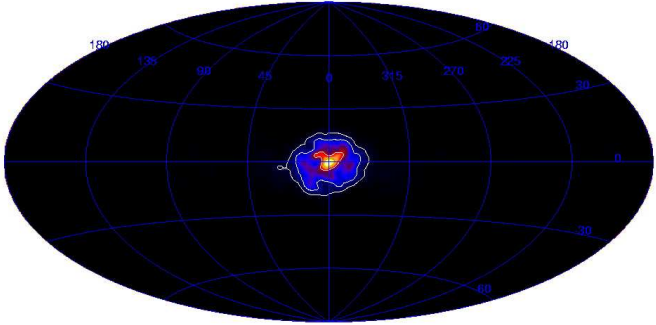, width=8.8cm}
  \end{center}
  \caption{First all-sky map of 511~keV electron-positron annihilation 
    radiation as observed by the SPI telescope aboard INTEGRAL
    (Kn\"odlseder et al.~2005).
    \label{fig:511keV}}
\end{figure}

The unprecedented imaging and spectroscopy capabilities of the 
spectrometer SPI aboard INTEGRAL have now provided for the first time 
an image of the distribution of 511~keV electron positron annihilation 
all over the sky (c.f.~Fig.~\ref{fig:511keV}; \cite{knoedl05}).
The outcome of this survey is astonishing: 511~keV line emission is 
only seen towards the bulge region of our Galaxy, while the rest of the 
sky remains surprisingly dark.
Only a weak glim of 511~keV emission is perceptible from the disk of 
the Galaxy, much less as expected from stellar populations following 
the global mass distribution of the Galaxy.
In other words, positron annihilation seems to be greatly enhanced in 
the bulge with respect to the disk of the Galaxy.

A detailed analysis of the 511~keV line shape measured by SPI has 
also provided interesting insights into the annihilation physics
(\cite{churazov05}).
At least two components have been identified, indicating that positron 
annihilation takes place in a partially ionised medium.
This clearly demonstrated that precise 511~keV line shape 
measurements provide important insights into the distribution of the 
various phases of the interstellar medium (ISM).

While INTEGRAL has set the global picture of galactic positron 
annihilation, high angular resolution mapping of the galactic bulge region 
is required to shed light on the still mysterious source of positrons.
So far, no individual source of positron emission could have been 
identified, primarily due to the expected low levels of 511~keV line 
fluxes.
An instrument with sufficiently good sensitivity and angular 
resolution should be able to pinpoint the origin of the positrons, 
by providing detailed maps of the central bulge region of the Galaxy.
With additional fine spectroscopic capabilities, comparable to that 
achieved by the germanium detectors onboard the SPI telescope, the spatial 
variations of the 511~keV line shape will allow to draw an unprecedented 
picture of the distribution of the various ISM phases in the inner regions 
of our Galaxy.

Thus, with the next generation gamma-ray telescope, galactic positrons 
will be exploited as a messenger from the mysterious antimatter source 
in the Milky-Way, as well as a tracer to probe the conditions of the 
ISM that are difficultly to measure by other means.

\section{Mission requirements}

The major mission requirement for the future European gamma-ray mission 
is sensitivity.
Many interesting scientific questions are in a domain where photons 
are rare (say $10^{-7}$ \funit), and therefore large collecting areas are 
needed to perform measurements in a reasonable amount of time.
It is clear that a significant sensitivity leap is required, say 
50--100 times more sensitive than current instruments, if the above 
listed scientific questions should be addressed.

With such a sensitivity leap, the expected number of observable sources 
would be large, implying the need for good angular resolution to avoid 
source confusion in\break crowded regions, such as for example the galactic 
centre.
Also, it is desirable to have an angular resolution comparable to that 
at other wavebands, to allow for source identification and hence multi 
wavelength studies.

As mentioned previously, gamma-ray emission may be substantially 
polarised due to the non-thermal nature of the underlying emission 
processes.
Studying not only the intensity but also the polarisation of the 
emission would add a new powerful scientific dimension to the 
observations.
Such measurements would allow to discriminate between the different 
plausible emission processes at work, and would allow to constrain the 
geometry of the emission sites.

Taking all these considerations into account, the following mission 
requirements derive (c.f.~Table \ref{tab:mission}).
The energy band should cover the soft gamma-ray band, with coverage 
down to the hard X-ray band (to overlap with future X-ray 
observatories), and coverage of the major gamma-ray lines of 
astrophysical interest.
A real sensitivity leap should be achieved, typical by a factor of 
50--100 with respect to existing gamma-ray instrumentation.
For high-resolution gamma-ray line spectroscopy a good energy resolution 
is desirable to exploit the full potential of line profile studies.
A reasonably sized field-of-view together with arcmin angular 
resolution should allow the imaging of field populations of gamma-ray 
sources in a single observation.
Finally, good polarisation capabilities, at the percent level for 
strong sources, are required to exploit this additional observable.

\begin{table}[bht]
  \caption{
    Mission requirements for the future European gamma-ray mission
    (sensitivities are for $10^6$ seconds at $3\sigma$ detection 
    significance).}
  \label{tab:mission}
  \begin{center}
    \leavevmode
    \footnotesize
    \begin{tabular}[h]{ll}
      \hline \\[-5pt]
      Parameter & Requirement \\[+5pt]
      \hline \\[-5pt]
      Energy band             & 50 keV -- 2 MeV \\
      Continuum sensitivity   & $10^{-8}$ \feunit\ \\
      Narrow line sensitivity & $5 \times 10^{-7}$ \funit\ \\
      Energy resolution       & 2 keV at 600 keV \\
      Field of view           & 30 arcmin \\
      Angular resolution      & arcmin \\
      Polarisation            & 1\% at 10 mCrab \\
      \hline \\
      \end{tabular}
  \end{center}
\end{table}

Can these mission requirements be reached within the 2015--2025 time frame?
We are convinced that the answer is yes.
How can these mission requirements be reached?
We think that the best solution is the implementation of a broad-band
gamma-ray lens telescope based on the principle of Laue diffraction of 
gamma-rays in mosaic crystals.
In the following section we explain why we come to this conclusion, and how 
such a future gamma-ray telescope for Europe may look like.

\section{A future gamma-ray telescope for Europe}

The sensitivity of current gamma-ray telescopes is\break severely
limited by the internal instrumental background and the size of the 
photon collecting area.
The instrumental background arises from the cosmic-ray bombardment of 
the telescope (and in particular with the gamma-ray detector material)
and the material around, and 
scales to first order with detector volume.\footnote{
  For energies $\la100$ keV a large fraction of the background is of
  astrophysical nature and arises from the isotropic cosmic gamma-ray 
  background. Therefore, the background at these energies depends 
  crucially on the solid angle covered by the field of view of the 
  instrument.}
For a given detector thickness, the instrumental background is 
therefore proportional to the geometrical detector area,
$C_{\rm B} \propto A_{\rm det}$.
The received signal scales proportionally to the collecting area,
$S \propto A_{\rm coll}$, and thus, the signal-to-noise ratio scales 
(assuming background domination) as
\begin{equation}
 S/N \approx \frac{S}{\sqrt{C_{\rm B}}} \propto 
 \frac{A_{\rm coll}}{\sqrt{A_{\rm det}}}
 \label{eq:sn}
\end{equation}

For conventional gamma-ray telescopes, such as the SPI and IBIS instruments 
aboard INTEGRAL, the detector area is identical to the collecting area,
i.e.~$A_{\rm det} = A_{\rm coll}$, and therefore the signal-to-noise 
ratio scales as
$S/N \propto \sqrt{A_{\rm coll}}$.
Therefore, if one would aim in increasing the INTEGRAL sensitivity by 
one order of magnitude, an instrument a hundred times bigger would be 
needed.
It is obvious that it is unrealistic to follow this way.
With INTEGRAL, the technique used for conventional gamma-ray 
telescope has probably reached its climax.

Decoupling the collecting area from the detection area, however, can 
change the situation dramatically.
For a collecting area comparable to that of the IBIS detector ISGRI
and a detection area as small as 1 cm$^2$, a sensitivity gain of 
approximately a factor of $\sim50$ would be achieved.
This would however mean that gamma-rays need to be focused or 
concentrated by a large area collector onto a small area detector, a 
technique that was for a long time believed to be inaccessible to 
gamma-ray astronomy.

Technological developments undertaken by several\break groups in Europe 
during the last years have now demonstrated that gamma-ray focusing is 
indeed possible (\cite{halloin04}; \cite{chiara00}).
The focusing is achieved by organising crystals onto rings or in 
an\break
Archimedes spiral around the optical axis, making use of small angle Bragg 
reflection in Laue geometry to deviate the incoming radiation onto a 
small focal spot.
Broad band energy coverage is achieved by using mosaic crystals with 
varying inclinations and radial distances to the optical axis
(\cite{ballmoos04};
Frontera et al., these proceedings;
von Ballmoos et al., these proceedings).
The gamma-ray lens principle has been demonstrated by various laboratory 
measurements, and also, during a successful balloon 
flight of the CLAIRE prototype telescope which led to the first ever 
detection of a gamma-ray source (the Crab nebula) by a focusing gamma-ray 
telescope (\cite{halloin04}).

The small diffraction angles of typically few tens of arcmin together 
with reasonable ring radii of the order of one meter or more, lead to 
substantial focal lengths of at least a few tens of meters, making it 
difficult to place the lens on the same satellite as the detector. 
Thus, a formation flight scenario comprising a lens spacecraft 
together with a detector spacecraft seems the most plausible 
configuration for such an instrument.
We note that such a gamma-ray lens telescope is currently under study 
at the French space agency CNES 
(project MAX; \cite{ballmoos04}) 
and at the ESA Science Payload \& Advanced Concepts Office
(project Gamma-Ray Lens),
which both confirm the feasibility of such a scenario.
We therefore believe that a gamma-ray lens telescope in formation 
flight configuration provides the most promising instrumental concept 
allowing advances in the field of space-based gamma-ray astronomy.

\begin{figure}[!t]
  \begin{center}
    \epsfig{file=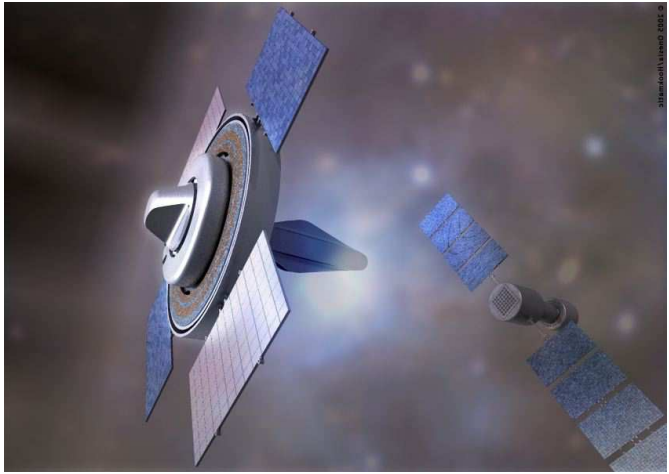, width=8.8cm}
  \end{center}
  \caption{Artists view of the future European gamma-ray telescope.
    A Laue lens, situated on the left spacecraft, is focusing 
    gamma-rays onto a small detector, situated on the right spacecraft.
    Both spacecrafts are in formation flight with a typical focal 
    length between a few tens and a few hundreds metres.
    \label{fig:lens}}
\end{figure}

The precise design of the gamma-ray lens telescope is currently under 
discussion in a dedicated working group.
The artists view in Fig.~\ref{fig:lens} may, however, give an idea 
how the future European gamma-ray telescope could look like.
In this example, the lens spacecraft is composed of concentric rings of 
crystals, where each ring is focusing a specific narrow energy band on the 
(same) focal spot on the detector spacecraft.
Higher energies show smaller diffraction angles and therefore are 
situated closer to the optical axis (inner rings).
Conversely, lower energies show larger diffraction angles and 
therefore are situated on the outer rings.
The lowest energies may require radial distances from the optical 
axis that exceed the available space in launcher fairings, therefore 
deployable lens petals may eventually be employed.

\section{Conclusions}

The gamma-ray band presents a unique astronomical window that allows the 
study of the most energetic and most violent phenomena in our Universe.
With ESA's INTEGRAL observatory, an unprecedented global survey of 
the soft gamma-ray sky is currently performed, revealing hundreds
of sources of different kinds, new classes of objects, extraordinary views 
of antimatter annihilation in our\break Galaxy, and fingerprints of recent 
nucleosynthesis processes.
While INTEGRAL provides the longly awaited\break global overview over the soft 
gamma-ray sky, there is a growing need to perform deeper, more 
focused investigations of gamma-ray sources, comparable to the 
step that has been taken in X-rays by going from the ROSAT survey 
satellite to the more focused XMM-Newton observatory.
Technological advances in the past years in the domain of gamma-ray 
focusing using Laue diffraction techniques have paved the way towards 
a future European gamma-ray mission, that will outreach past missions 
by large factors in sensitivity and angular resolution.
Such a future {\em Gamma-Ray Imager} will allow to study particle 
acceleration processes and explosion physics in unprecedented depth, 
providing essential clues on the intimate nature of the most violent 
and most energetic processes in the Universe.

\begin{acknowledgements}

J. Kn\"odlseder would like to thank the European gamma-ray community for 
its enthusiasm in preparing the future.
In particular he thanks the Gamma-Ray Imager Science Working Group 
that has very actively participated in preparing the presentation at 
the ESLAB symposium and in writing the present paper.

\end{acknowledgements}


\end{document}